\documentclass[aps,prl,twocolumn,showpacs,superscriptaddress,groupedaddress]{revtex4}
\usepackage{graphicx}  
\usepackage[usenames]{color} 
\usepackage{dcolumn}   
\usepackage{bm}        
\usepackage{amssymb,amsmath}   

\newcommand{\pt}        {\mbox{$p_T$}}
\newcommand{\met}       {\mbox{\ensuremath{\slash\kern-.7emE_{T}}}}

\newcommand{\rargap}    {\mbox{ $\rightarrow$ }}

\newcommand{\ttbar}     {\mbox{$t\bar{t}$}}

\newcommand{\ppbar}     {\mbox{$p\bar{p}$}}

\newcommand{\singletop} {\sc{singletop}}

\newcommand{\pythia}    {\sc{pythia}}
\newcommand{\alpgen}    {\sc{alpgen}}

\newcommand{\geant}     {\sc{geant}}

\newcommand{\SM}{standard model}

\begin{document}
\hspace{5.2in} \mbox{FERMILAB-PUB-11-561-E-PPD}

\title{Search for anomalous $\boldsymbol{Wtb}$ couplings in single top quark production in $\boldsymbol{\ppbar}$ collisions at $\boldsymbol{\sqrt{s}=1.96}$~TeV}
\affiliation{Universidad de Buenos Aires, Buenos Aires, Argentina}
\affiliation{LAFEX, Centro Brasileiro de Pesquisas F{\'\i}sicas, Rio de Janeiro, Brazil}
\affiliation{Universidade do Estado do Rio de Janeiro, Rio de Janeiro, Brazil}
\affiliation{Universidade Federal do ABC, Santo Andr\'e, Brazil}
\affiliation{Instituto de F\'{\i}sica Te\'orica, Universidade Estadual Paulista, S\~ao Paulo, Brazil}
\affiliation{University of Science and Technology of China, Hefei, People's Republic of China}
\affiliation{Universidad de los Andes, Bogot\'{a}, Colombia}
\affiliation{Charles University, Faculty of Mathematics and Physics, Center for Particle Physics, Prague, Czech Republic}
\affiliation{Czech Technical University in Prague, Prague, Czech Republic}
\affiliation{Center for Particle Physics, Institute of Physics, Academy of Sciences of the Czech Republic, Prague, Czech Republic}
\affiliation{Universidad San Francisco de Quito, Quito, Ecuador}
\affiliation{LPC, Universit\'e Blaise Pascal, CNRS/IN2P3, Clermont, France}
\affiliation{LPSC, Universit\'e Joseph Fourier Grenoble 1, CNRS/IN2P3, Institut National Polytechnique de Grenoble, Grenoble, France}
\affiliation{CPPM, Aix-Marseille Universit\'e, CNRS/IN2P3, Marseille, France}
\affiliation{LAL, Universit\'e Paris-Sud, CNRS/IN2P3, Orsay, France}
\affiliation{LPNHE, Universit\'es Paris VI and VII, CNRS/IN2P3, Paris, France}
\affiliation{CEA, Irfu, SPP, Saclay, France}
\affiliation{IPHC, Universit\'e de Strasbourg, CNRS/IN2P3, Strasbourg, France}
\affiliation{IPNL, Universit\'e Lyon 1, CNRS/IN2P3, Villeurbanne, France and Universit\'e de Lyon, Lyon, France}
\affiliation{III. Physikalisches Institut A, RWTH Aachen University, Aachen, Germany}
\affiliation{Physikalisches Institut, Universit{\"a}t Freiburg, Freiburg, Germany}
\affiliation{II. Physikalisches Institut, Georg-August-Universit{\"a}t G\"ottingen, G\"ottingen, Germany}
\affiliation{Institut f{\"u}r Physik, Universit{\"a}t Mainz, Mainz, Germany}
\affiliation{Ludwig-Maximilians-Universit{\"a}t M{\"u}nchen, M{\"u}nchen, Germany}
\affiliation{Fachbereich Physik, Bergische Universit{\"a}t Wuppertal, Wuppertal, Germany}
\affiliation{Panjab University, Chandigarh, India}
\affiliation{Delhi University, Delhi, India}
\affiliation{Tata Institute of Fundamental Research, Mumbai, India}
\affiliation{University College Dublin, Dublin, Ireland}
\affiliation{Korea Detector Laboratory, Korea University, Seoul, Korea}
\affiliation{CINVESTAV, Mexico City, Mexico}
\affiliation{Nikhef, Science Park, Amsterdam, the Netherlands}
\affiliation{Radboud University Nijmegen, Nijmegen, the Netherlands and Nikhef, Science Park, Amsterdam, the Netherlands}
\affiliation{Joint Institute for Nuclear Research, Dubna, Russia}
\affiliation{Institute for Theoretical and Experimental Physics, Moscow, Russia}
\affiliation{Moscow State University, Moscow, Russia}
\affiliation{Institute for High Energy Physics, Protvino, Russia}
\affiliation{Petersburg Nuclear Physics Institute, St. Petersburg, Russia}
\affiliation{Instituci\'{o} Catalana de Recerca i Estudis Avan\c{c}ats (ICREA) and Institut de F\'{i}sica d'Altes Energies (IFAE), Barcelona, Spain}
\affiliation{Stockholm University, Stockholm and Uppsala University, Uppsala, Sweden}
\affiliation{Lancaster University, Lancaster LA1 4YB, United Kingdom}
\affiliation{Imperial College London, London SW7 2AZ, United Kingdom}
\affiliation{The University of Manchester, Manchester M13 9PL, United Kingdom}
\affiliation{University of Arizona, Tucson, Arizona 85721, USA}
\affiliation{University of California Riverside, Riverside, California 92521, USA}
\affiliation{Florida State University, Tallahassee, Florida 32306, USA}
\affiliation{Fermi National Accelerator Laboratory, Batavia, Illinois 60510, USA}
\affiliation{University of Illinois at Chicago, Chicago, Illinois 60607, USA}
\affiliation{Northern Illinois University, DeKalb, Illinois 60115, USA}
\affiliation{Northwestern University, Evanston, Illinois 60208, USA}
\affiliation{Indiana University, Bloomington, Indiana 47405, USA}
\affiliation{Purdue University Calumet, Hammond, Indiana 46323, USA}
\affiliation{University of Notre Dame, Notre Dame, Indiana 46556, USA}
\affiliation{Iowa State University, Ames, Iowa 50011, USA}
\affiliation{University of Kansas, Lawrence, Kansas 66045, USA}
\affiliation{Kansas State University, Manhattan, Kansas 66506, USA}
\affiliation{Louisiana Tech University, Ruston, Louisiana 71272, USA}
\affiliation{Boston University, Boston, Massachusetts 02215, USA}
\affiliation{Northeastern University, Boston, Massachusetts 02115, USA}
\affiliation{University of Michigan, Ann Arbor, Michigan 48109, USA}
\affiliation{Michigan State University, East Lansing, Michigan 48824, USA}
\affiliation{University of Mississippi, University, Mississippi 38677, USA}
\affiliation{University of Nebraska, Lincoln, Nebraska 68588, USA}
\affiliation{Rutgers University, Piscataway, New Jersey 08855, USA}
\affiliation{Princeton University, Princeton, New Jersey 08544, USA}
\affiliation{State University of New York, Buffalo, New York 14260, USA}
\affiliation{Columbia University, New York, New York 10027, USA}
\affiliation{University of Rochester, Rochester, New York 14627, USA}
\affiliation{State University of New York, Stony Brook, New York 11794, USA}
\affiliation{Brookhaven National Laboratory, Upton, New York 11973, USA}
\affiliation{Langston University, Langston, Oklahoma 73050, USA}
\affiliation{University of Oklahoma, Norman, Oklahoma 73019, USA}
\affiliation{Oklahoma State University, Stillwater, Oklahoma 74078, USA}
\affiliation{Brown University, Providence, Rhode Island 02912, USA}
\affiliation{University of Texas, Arlington, Texas 76019, USA}
\affiliation{Southern Methodist University, Dallas, Texas 75275, USA}
\affiliation{Rice University, Houston, Texas 77005, USA}
\affiliation{University of Virginia, Charlottesville, Virginia 22901, USA}
\affiliation{University of Washington, Seattle, Washington 98195, USA}
\author{V.M.~Abazov} \affiliation{Joint Institute for Nuclear Research, Dubna, Russia}
\author{B.~Abbott} \affiliation{University of Oklahoma, Norman, Oklahoma 73019, USA}
\author{B.S.~Acharya} \affiliation{Tata Institute of Fundamental Research, Mumbai, India}
\author{M.~Adams} \affiliation{University of Illinois at Chicago, Chicago, Illinois 60607, USA}
\author{T.~Adams} \affiliation{Florida State University, Tallahassee, Florida 32306, USA}
\author{G.D.~Alexeev} \affiliation{Joint Institute for Nuclear Research, Dubna, Russia}
\author{G.~Alkhazov} \affiliation{Petersburg Nuclear Physics Institute, St. Petersburg, Russia}
\author{A.~Alton$^{a}$} \affiliation{University of Michigan, Ann Arbor, Michigan 48109, USA}
\author{G.~Alverson} \affiliation{Northeastern University, Boston, Massachusetts 02115, USA}
\author{G.A.~Alves} \affiliation{LAFEX, Centro Brasileiro de Pesquisas F{\'\i}sicas, Rio de Janeiro, Brazil}
% ...
%
%
\collaboration{The D0 Collaboration\footnote{with visitors from
$^{a}$Augustana College, Sioux Falls, SD, USA,
$^{b}$The University of Liverpool, Liverpool, UK,
$^{c}$UPIITA-IPN, Mexico City, Mexico,
$^{d}$SLAC, Menlo Park, CA, USA,
$^{e}$University College London, London, UK,
$^{f}$Centro de Investigacion en Computacion - IPN, Mexico City, Mexico,
$^{g}$ECFM, Universidad Autonoma de Sinaloa, Culiac\'an, Mexico,
and 
$^{h}$Universit{\"a}t Bern, Bern, Switzerland.
$^{\ddag}$Deceased.
}} \noaffiliation
\vskip 0.25cm

\date{October 19, 2011}

\begin{abstract}

We present new direct constraints on a general $Wtb$ interaction using data corresponding to
an integrated luminosity of 5.4~fb$^{-1}$ collected by the D0~detector at the Tevatron {\ppbar} collider. 
The standard model provides a purely left-handed vector coupling
at the $Wtb$ vertex, while the most general, lowest dimension
Lagrangian allows right-handed vector and left- or right-handed tensor
couplings as well. 
We obtain precise limits on these anomalous couplings by
comparing the data to the expectations from different assumptions on the $Wtb$ coupling.
\end{abstract}

\maketitle

\newpage

\begin{figure}
\vspace{0.1in}
\includegraphics[width=0.45\textwidth]{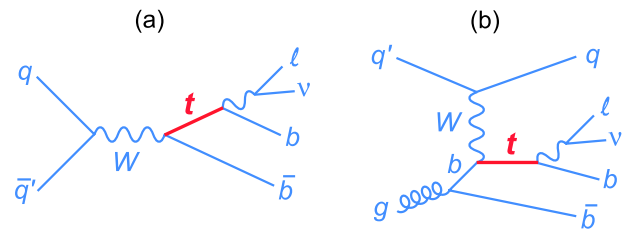}
\vspace{-0.1in}
\caption{Tree level Feynman diagrams for (a) $tb$ and (b) $tqb$ single top quark production.}
\label{fig:feynman}
\end{figure}

The top quark was discovered in 1995 at the Tevatron~\cite{top-obs-1995-d0,top-obs-1995-cdf} 
via the pair production mode involving strong interactions. 
In 2009, the electroweak production of the top quark was observed by 
the D0 and CDF collaborations~\cite{stop-obs-2009-d0,stop-obs-2009-cdf}. At the Tevatron, the dominant production 
modes for single top quark are the $s$-channel (``$tb$") 
~\cite{singletop-cortese} and $t$-channel (``$tqb$")
~\cite{singletop-willenbrock,singletop-yuan} processes illustrated in Figure~\ref{fig:feynman}.
A third process, usually called ``associated production'' in which the top
quark is produced together with a W boson, has a small cross section at the
Tevatron compared to the $tb$ and $tqb$ processes~\cite{singletop-xsec-kidonakis}. 
Recently, we presented improved measurements of the single top quark production 
cross sections~\cite{st-channel-new} and the observation of $t$-channel single top quark production~\cite{t-channel-new}.

The large mass of the top quark implies that it has large couplings to the electroweak symmetry 
breaking sector of the \SM~(SM) and may have non-standard interactions with the weak gauge bosons.
Single top quark production provides a unique probe to study the interactions of the top quark with
the $W$ boson.

The most general, lowest dimension, $CP$-conserving {\it Wtb} vertex is given by~\cite{lagrangian}:

\begin{equation}
\begin{split}
\mathcal{L}& = -\frac{g}{\sqrt{2}}\bar{b} \gamma^\mu
(L_V P_L + R_V P_R) t W_{\mu}^{-} \\
&-\frac{g}{\sqrt{2}} \bar{b} \frac{i\sigma^{\mu\nu} q_{\nu}}{M_W} 
(L_T P_L + R_T P_R) t W_{\mu}^{-} + h.c. ,
\end{split}
\label{coupling}
\end{equation}
\noindent where $M_W$ is the mass of the $W$ boson, $q_{\nu}$ is the {\it{W}} boson four-momentum, 
$P_{L}=(1 - \gamma_5)/2$ is the left-handed projection operator, $P_{R}=(1 + \gamma_5)/2$ is the 
right-handed projection operator, $L_{V,T}= V_{tb} \cdot f_{L_{V,T}}$ and $R_{V,T} = V_{tb} \cdot f_{R_{V,T}}$.
The form factor $f_{L_{V}}$ ($f_{L_{T}}$) represents the left-handed vector (tensor) coupling, $f_{R_{V}}$ ($f_{R_{T}}$) represents the right-handed vector 
(tensor) coupling, and $V_{tb}$ is the Cabibbo-Kobayashi-Maskawa matrix element. 
In the SM, the $Wtb$ coupling is left-handed with $L_V \equiv |V_{tb}| \simeq 1 
$ and $R_V=L_T=R_T =0$.
The magnitudes of the right-handed vector coupling and the tensor couplings can be indirectly constrained by the 
measured branching ratio of the $b \rightarrow s\gamma$ process~\cite{bsgamma}.
Measurements of top quark decays in $t\bar{t}$ production, e.g. the $W$  
boson helicity~\cite{d0-whelicity,cdf-whelicity}, can directly constrain the Lorentz structure of the {\it Wtb} 
vertex~\cite{anom-prl-2009}.
Assuming single top quarks are produced only via {\it{W}} boson exchange, the single top quark cross section is 
directly proportional to the square of the effective {\it Wtb} coupling. Moreover, the event kinematics and angular 
distributions are also sensitive to the existence of anomalous top quark couplings~\cite{dudko-boos,Wtb-coupling-yuan-etal}. 
Therefore, direct constraints on anomalous couplings can be obtained by measuring single top quark 
production~\cite{anom-prl-2008}.

This analysis uses the same data, event selection, and background modeling as the recent single top quark cross section 
measurements~\cite{t-channel-new, st-channel-new}.
We perform a study of anomalous $Wtb$ couplings and obtain substantial improvements on the limits of these 
couplings following the general framework given in Ref.~\cite{anom-prl-2008}. 
Out of the four couplings ($L_V$, $L_T$, $R_V$, $R_T$), we consider three cases pairing the left-handed vector 
coupling with each of the other three couplings: ($L_V$, $R_V$), ($L_V$, $L_T$) and ($L_V$, $R_T$), and for each case 
we assume the other two non-SM couplings are negligible. This pairing allows us to 
limit the complexity of the analysis, while increasing the statistical power and sensitivity for the
anomalous coupling under study. 
We assume that single top quarks are produced exclusively through $W$ boson exchange. Therefore other single top quark 
production mechanisms, such as flavor-changing neutral current interactions~\cite{d0-fcnc,cdf-fcnc}, the decay of new 
scalar boson~\cite{d0-chargehiggs}, or the exchange of new vector boson~\cite{d0-wprime,cdf-wprime} are not considered here.
We also assume that the $Wtb$ vertex dominates top quark production and decay, i.e., 
$|V_{td}|^2+|V_{ts}|^2 \ll |V_{tb}|^2$. 
The results presented here supersede those contained in Ref.~\cite{anom-prl-2008}. 
In addition to the analysis of additional data and to the improvements in the event selection and background modeling 
used for the results presented in Ref.~\cite{t-channel-new, st-channel-new}, we use an updated calculation of the 
cross section as a function of the anomalous couplings which addresses a mistake present in the original 
analysis~\cite{anom-prl-2008} (in that analysis the $R_T$ and $L_T$ couplings were accidentally swapped).

We select single top quark events which are expected to contain exactly one isolated large transverse momentum ($\pt$) 
electron or muon and large missing transverse energy ($\met$). Events with 2, 3 or 4 jets are selected, and one or two of the jets are required to originate 
from the hadronization of long-lived $b$ hadrons ($b$-jets) as determined by a multivariate $b$-tagging algorithm~\cite{D0btag}. 
To increase the search sensitivity, we divide our data into six independent analysis channels, each with a different background composition 
and signal-to-background ratio. The channels are based on the number of identified $b$ jets (1 or 2)
and jet multiplicity (2, 3 or 4 jets). 
The signal selection efficiencies with different $Wtb$ couplings, including branching fraction, trigger efficiencies 
and the $b$-tagging requirements, vary between $2.7$\% and $3.0$\% for $tb$ and $1.9$\% 
and $2.2$\% for $tqb$ production, estimated using Monte Carlo (MC) simulations. The ``associated production'' process gives
a negligible contribution to this analysis.

Single top quark signal events with the SM and anomalous $Wtb$ couplings are modeled 
using the {\sc comphep}-based MC event generator {\singletop}~\cite{singletop-mcgen}
for a top quark mass $m_t = 172.5$~GeV using the CTEQ6M~\cite{cteq} parton distribution functions.
The anomalous $Wtb$ couplings are taken into account in both production and decay in the generated samples. 
The event kinematics for both $s$-channel and $t$-channel processes reproduce distributions from next-to-leading-order 
calculations~\cite{singletop-xsec-sullivan, Campbell:2009ss}. The decay of the top quark and the resulting $W$~boson 
are carried out in the {\sc singletop}~\cite{singletop-mcgen} generator in order to preserve information about 
the spin of the particles. The theoretical cross sections for anomalous single top quark production 
($s$+$t$-channel) with $|V_{tb}| \simeq 1$ are $3.1\pm0.3$~pb if $f_{R_V}=1$, $9.4\pm1.4$~pb if $f_{L_T}=1$ or $f_{R_T}=1$,
and $10.6\pm0.8$~pb if $f_{R_T}=f_{L_V}=1$~\cite{dudko-boos}. All other couplings are set to zero when 
calculating these cross sections. 
The SM single top quark production cross section is $3.3\pm0.1$~pb~\cite{singletop-xsec-kidonakis}.

The main background contributions are those from $W$~bosons
produced in association with jets ($W$+jets), {\ttbar} production, and multijet
production in which a jet with high electromagnetic
content mimics an electron, or a muon contained within a jet originating from
the decay of a heavy-flavor quark ($b$ or $c$ quark) appears to be isolated.
Diboson ($WW$, $WZ$, $ZZ$) and  $Z$+jets processes
add small additional contributions to the background.
The {\ttbar}, $W$+jets, and $Z$+jets events are simulated with the {\alpgen} leading-log MC
generator~\cite{alpgen}. The effect of anomalous $Wtb$ couplings on kinematic distributions of 
the {\ttbar} background has been found to be negligible, thus only SM {\ttbar} samples are considered 
in the analysis. Diboson processes are modeled using {\pythia}~\cite{pythia}. For all of the signal and background MC
samples, {\pythia} is used to simulate parton showers and to model hadronization of all 
generated partons. The presence of additional {\ppbar} interactions is modeled
by events selected from random beam crossings matching the instantaneous luminosity profile in the data.
All MC events are processed through a {\geant}-based simulation~\cite{geant} of the D0 detector and 
reconstructed using the same algorithm as data. Differences between simulation and data in lepton and jet reconstruction efficiencies and resolutions, jet
energy scale, and $b$-tagging efficiencies are corrected in the simulation by applying correction functions measured
from separate data samples. The {\ttbar}, $Z$+jets and diboson MC samples are scaled to their theoretical cross 
sections~\cite{ttbar-xsec,mcfm}. We use data containing non-isolated leptons to model the multijet
background. $W$+jets and multijet backgrounds are normalized by comparing the prediction for
background to data before $b$-tagging. 
Details of the selection criteria and background modeling are given in Ref.~\cite{st-channel-new}.

The main contributions to the systematic uncertainty on the predicted number of events arise from the signal modeling,
the jet energy scale (JES), jet energy resolution (JER), corrections to $b$-tagging efficiency and the correction for jet-flavor 
composition in $W$+jets events. These uncertainties affect the normalization of the distributions, and in some cases 
(JES, JER, and $b$-tagging) also change the differential distributions.
There are smaller contributions due to limited statistics of the MC samples, 
uncertainties on the measured luminosity, and the trigger modeling. In addition, we also consider a signal cross 
section uncertainty ($3.8$\% for $tb$ and $5.3$\% for $tqb$) given by the NLO calculation. 
Details of systematic uncertainties are given in Ref.~\cite{st-channel-new}. 
Table~\ref{tab:event-yields} lists the numbers of events expected and observed for each process as a function of jet multiplicity.

\begin{table}[!h!tbp]
\vspace{-0.15in}
\begin{center}
\caption{Numbers of expected and observed events in $5.4\;\rm fb^{-1}$ of integrated luminosity,
with uncertainties including both statistical and systematic components. The single top quark contributions are
normalized to their theoretical predictions.
}
\label{tab:event-yields}
\begin{tabular}{lrclrclrcl}
\hline\hline
Source     & \multicolumn{3}{c}{2 jets} & \multicolumn{3}{c}{3 jets} & \multicolumn{3}{c}{4 jets} \\
\hline
$tb\ $  ($f_{L_T}=1$)       & 756 &$\pm$& 42 & 344 &$\pm$& 27 & 103 &$\pm$& 15 \\
$tqb$   ($f_{L_T}=1$)       & 103 &$\pm$& 5.8 & 67 &$\pm$& 6.3 & 28 &$\pm$& 4.4 \\
$tb\ $  ($f_{R_V}=1$)        & 105 &$\pm$& 6.0 & 43 &$\pm$& 3.8 & 12 &$\pm$& 1.9 \\
$tqb$   ($f_{R_V}=1$)       & 122 &$\pm$& 7.2 & 61 &$\pm$& 5.3 & 22 &$\pm$& 3.7 \\
$tb\ $  ($f_{R_T}=1$)      & 730 &$\pm$& 38 & 316 &$\pm$& 25 & 92 &$\pm$& 14 \\
$tqb$   ($f_{R_T}=1$)      & 117 &$\pm$& 6.2 & 86 &$\pm$& 8.6 & 40 &$\pm$& 5.8 \\
$tb\ $  ($f_{L_V}=f_{R_T}=1$)      & 607 &$\pm$& 31 & 284 &$\pm$& 21 & 86 &$\pm$& 13 \\
$tqb$   ($f_{L_V}=f_{R_T}=1$)     & 268 &$\pm$& 15 & 167 &$\pm$& 16 & 67 &$\pm$& 10 \\
\hline
$tb\ $  (SM, $f_{L_V}=1$)       & 104 &$\pm$& 16 & 44 &$\pm$& 7.8 & 13 &$\pm$& 3.5 \\
$tqb$   (SM, $f_{L_V}=1$)      & 140 &$\pm$& 13 & 72 &$\pm$& 9.4 & 26 &$\pm$& 6.4 \\
${\ttbar}$ & 433 &$\pm$& 87 & 830 &$\pm$& 133 & 860 &$\pm$& 163 \\
$W$+jets   & 3,560 &$\pm$& 354 & 1,099 &$\pm$& 169 & 284 &$\pm$& 76 \\
$Z$+jets and dibosons & 400 &$\pm$& 55 & 142 &$\pm$& 41 & 35 &$\pm$& 18 \\
Multijets  & 277 &$\pm$& 34 & 130 &$\pm$& 17 & 43 &$\pm$& 5.2 \\
\hline
Total SM prediction & 4,914 &$\pm$& 558 & 2,317 &$\pm$& 377 & 1,261 &$\pm$& 272 \\
\hline
Data       & \multicolumn{3}{c}{4,881} & \multicolumn{3}{c}{2,307} & \multicolumn{3}{c}{1,283} \\
\hline\hline
\end{tabular}
\end{center}
\vspace{-0.15in}
\end{table}

\begin{figure*}[!h!tbp]
\includegraphics[width=0.30\textwidth]{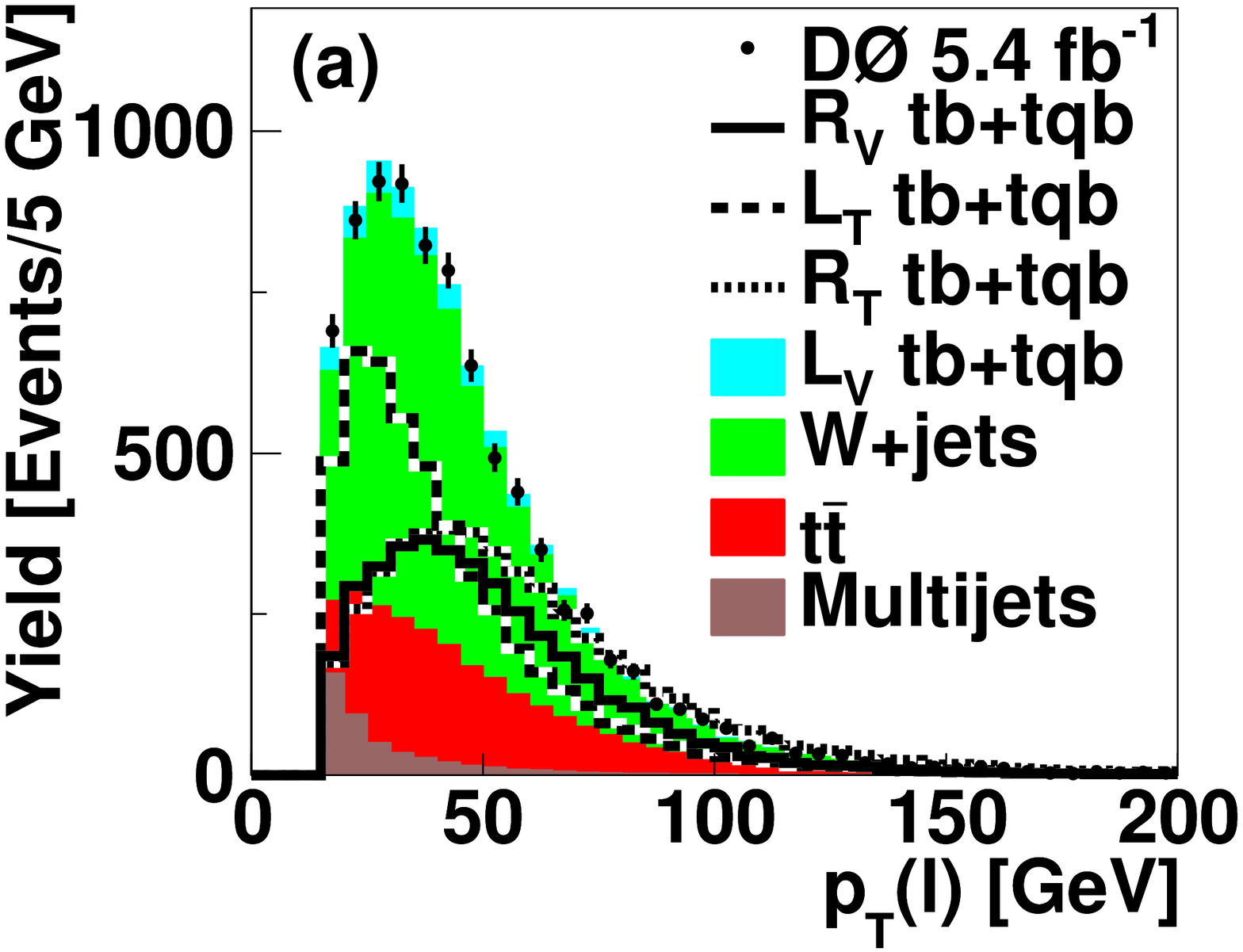}
\includegraphics[width=0.30\textwidth]{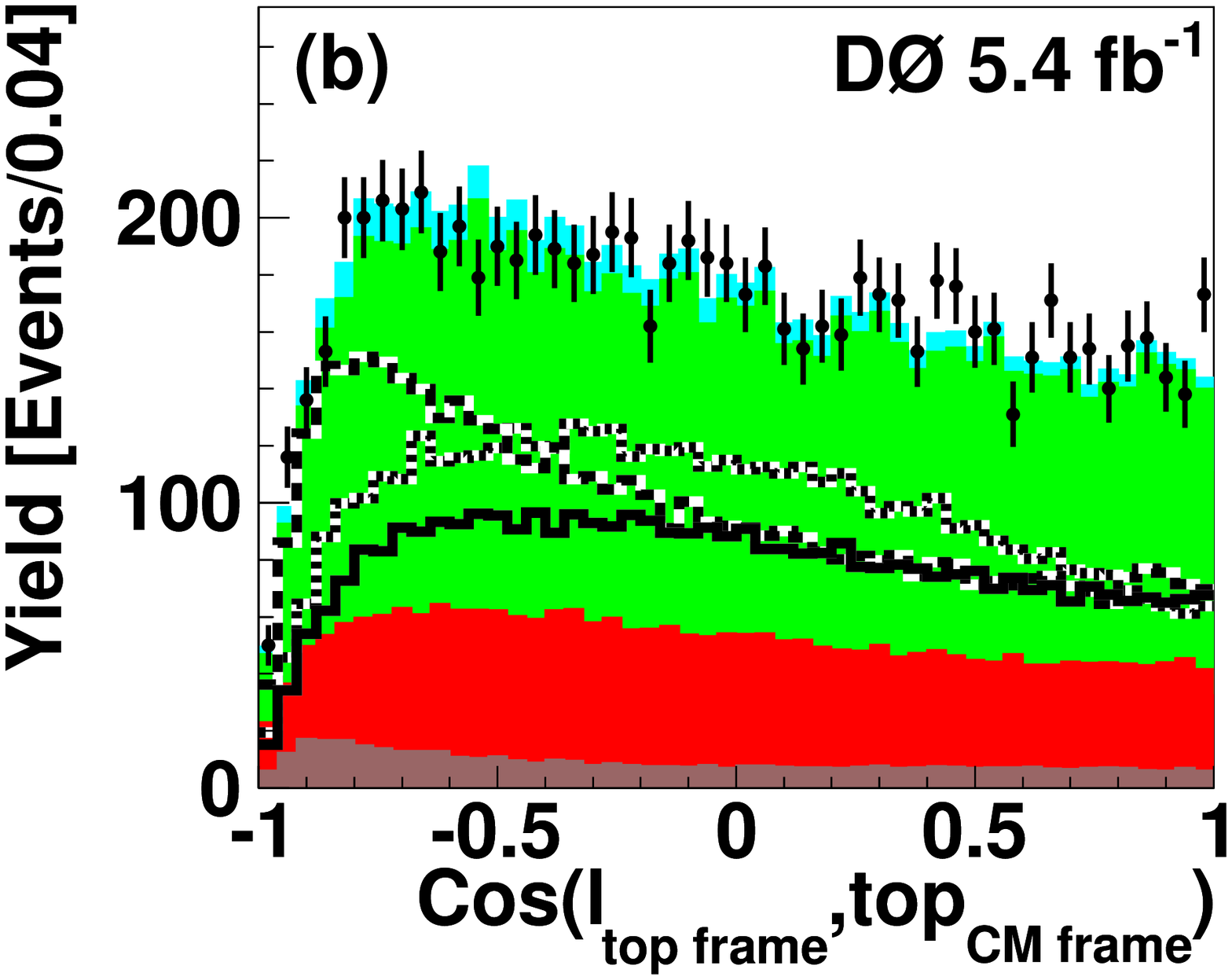}
\includegraphics[width=0.30\textwidth]{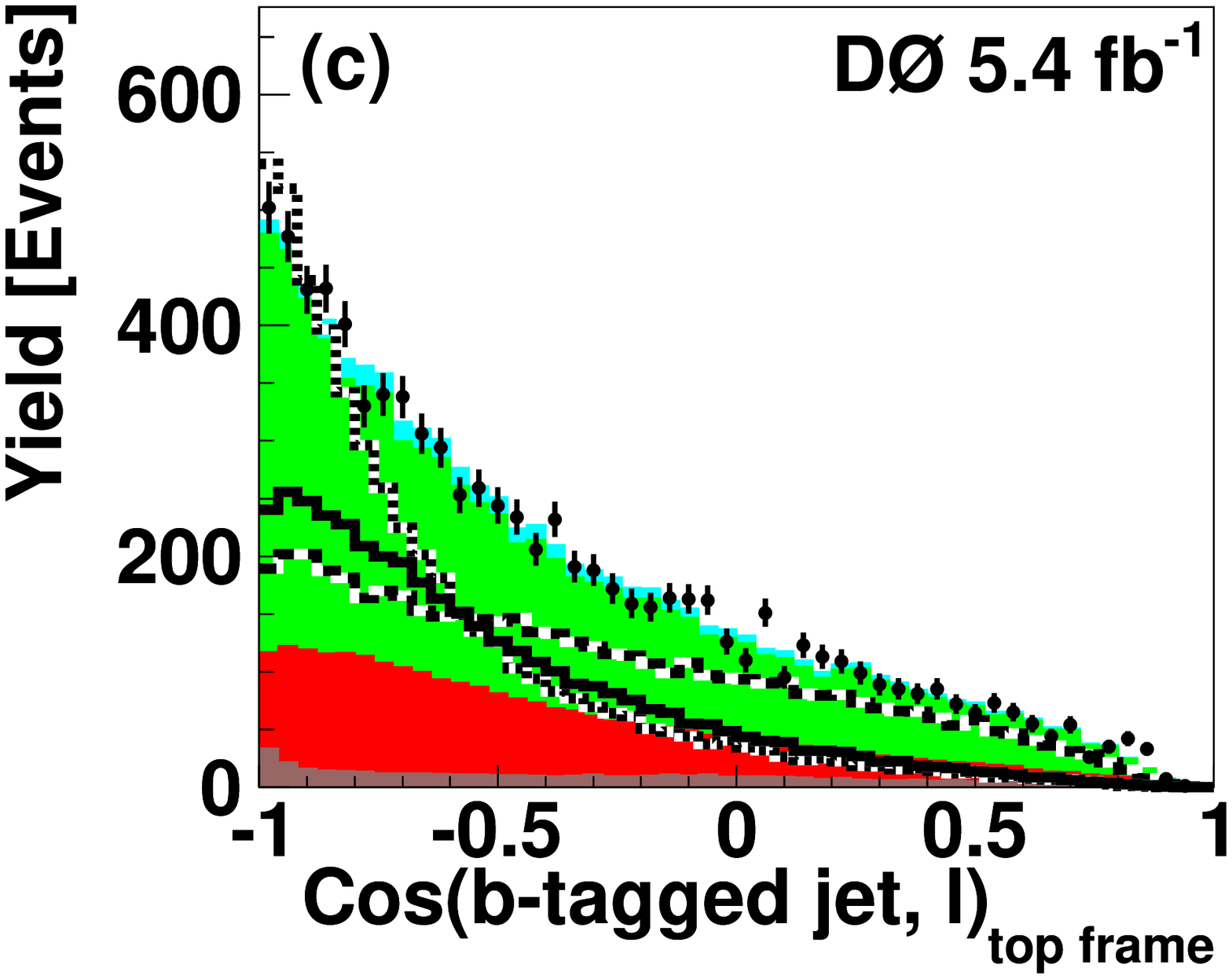}
\vspace{-0.1in}
\caption[variableplots]{
Comparison of the SM backgrounds and data for selected discriminating variables with all channels combined: (a) lepton $p_T$, 
(b) cosine of the angle between the lepton (in the reconstructed top quark frame) and the reconstructed top quark (in the center of mass frame) 
and (c) cosine of the angle between the leading $b$-tagged jet and the lepton (both in the reconstructed top quark frame).
Superimposed are the distributions from single top quark production (``$tb+tqb$") with 
one non-vanishing non-SM coupling (all other couplings set to zero) 
normalized to 10 times the SM single top quark cross section. The $W$+jets contributions include the smaller
backgrounds from $Z$+jets and dibosons.}
\label{variableplots}
\vspace{-.1in}
\end{figure*}

We use a multivariate analysis technique called Bayesian neural networks (BNN)~\cite{bayesianNNs} 
to separate the signal from the backgrounds.
The BNN discriminant is trained using the lepton and jets four-vectors, a two-vector for $\met$, and variables 
that include lepton charge and $b$-tagging information. In addition, 
four angular variables are added based on top quark spin
and $W$ boson helicity information to provide additional discriminating power. The total number of variables
used in training for events with 2, 3, and 4 jets is 18, 22, and 26, respectively~\cite{st-channel-new}. 
Three example distributions from some of the most sensitive variables are shown in 
Figure~\ref{variableplots}: the $p_T$ spectrum of the lepton from
the decay of the top quark and the cosine of the angles between the lepton, the leading $b$-tagged jet and the reconstructed top quark.

\begin{figure}[!h!tbp]
\includegraphics[width=0.235\textwidth]{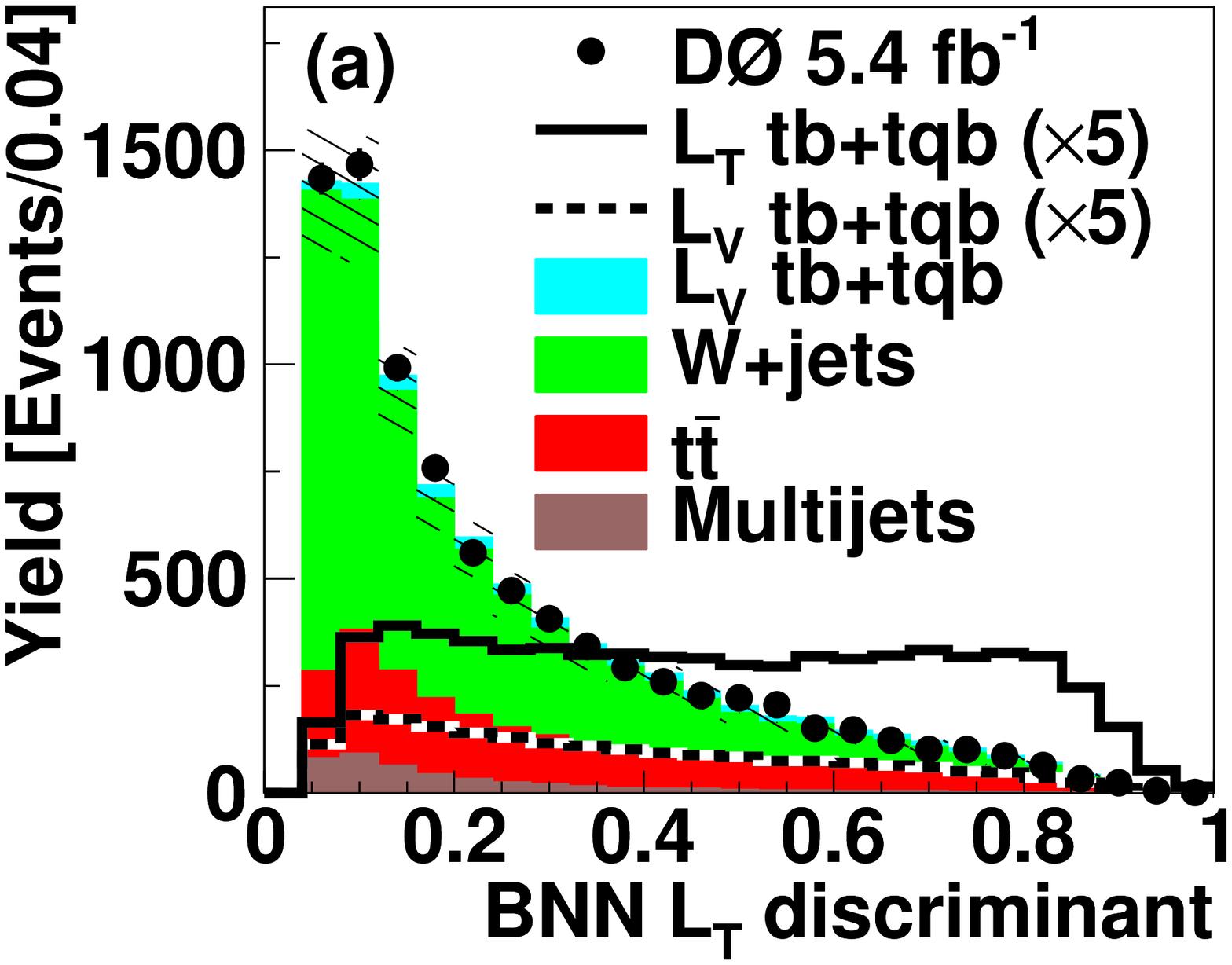}
\includegraphics[width=0.235\textwidth]{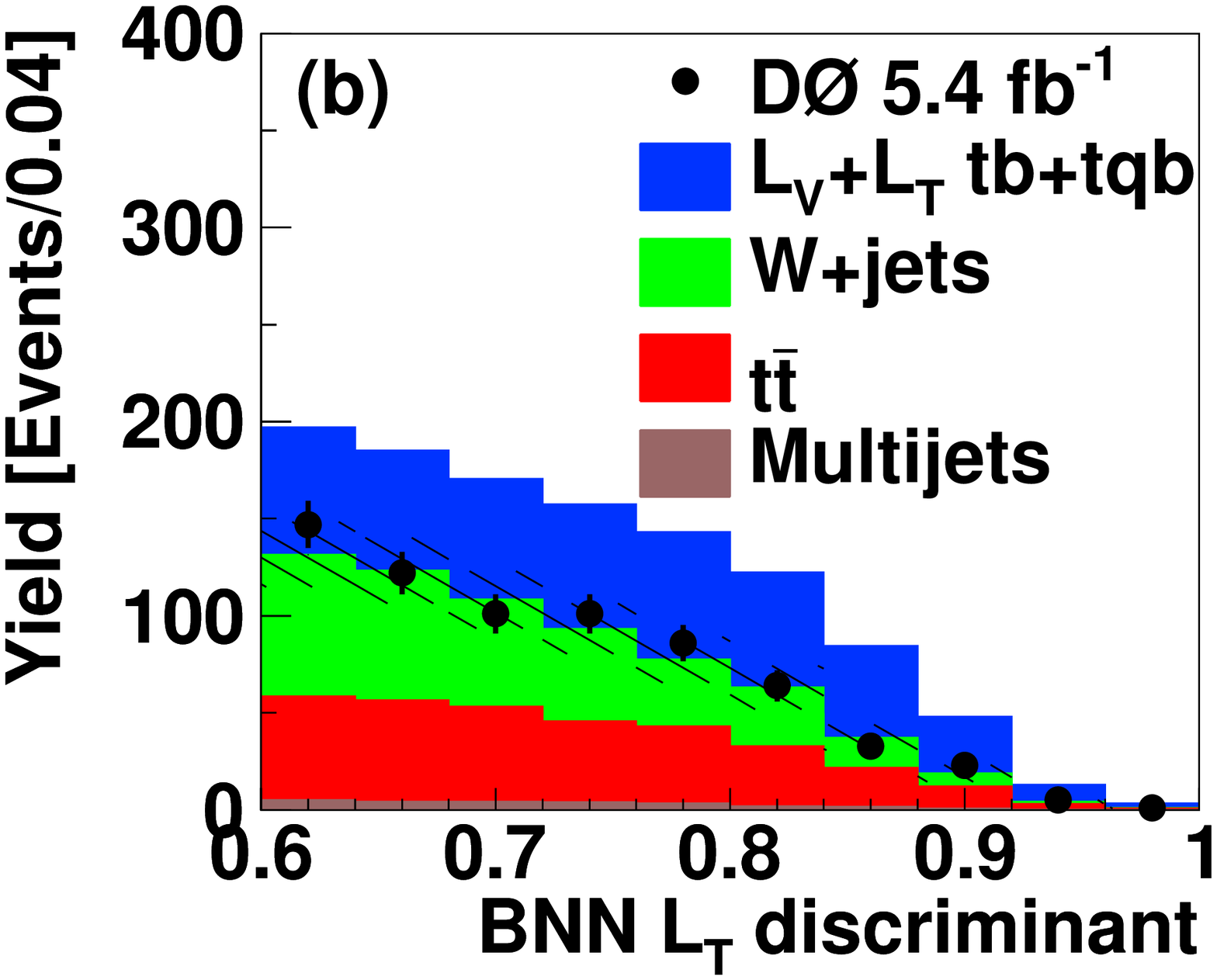}
\includegraphics[width=0.235\textwidth]{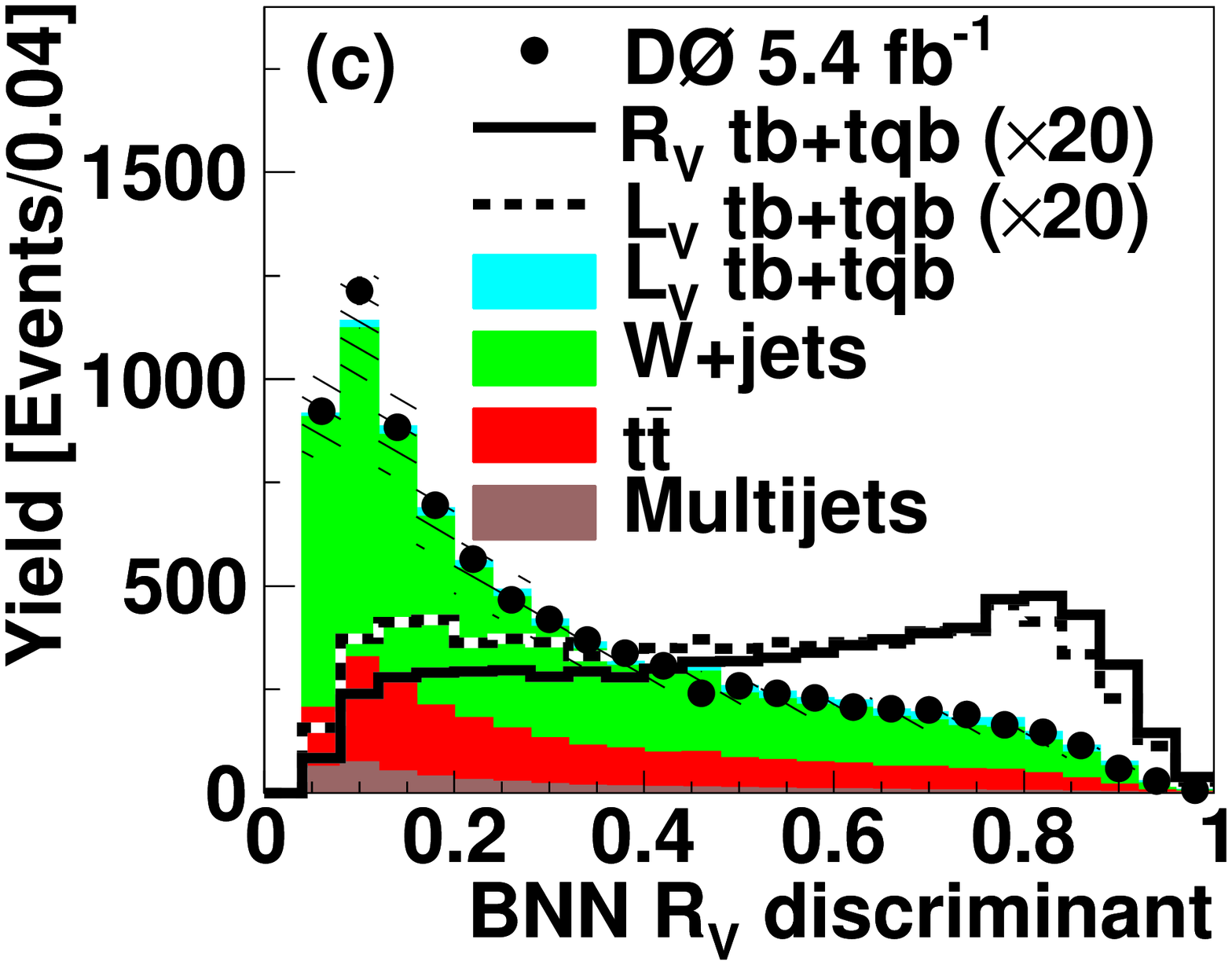}
\includegraphics[width=0.235\textwidth]{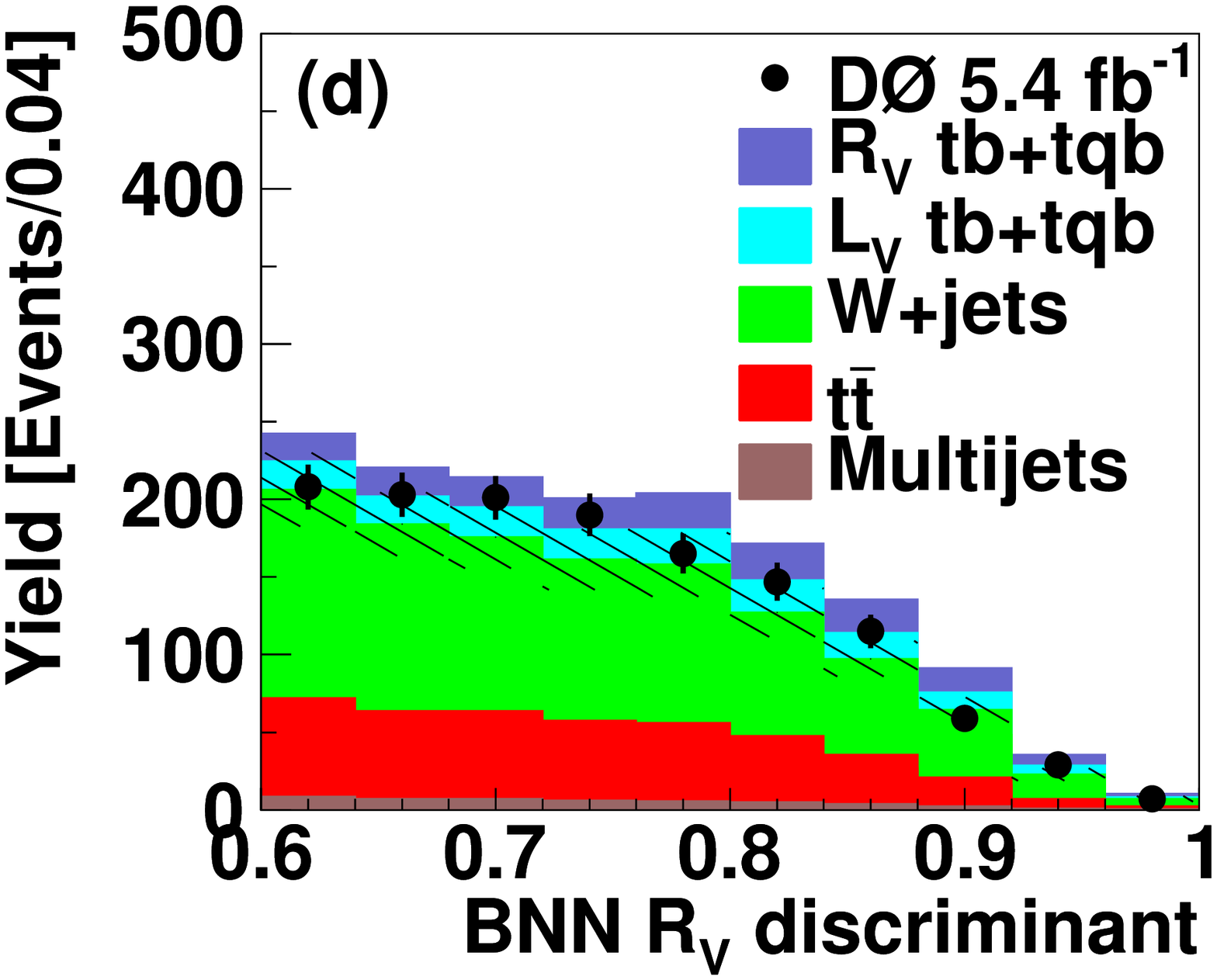}
\includegraphics[width=0.235\textwidth]{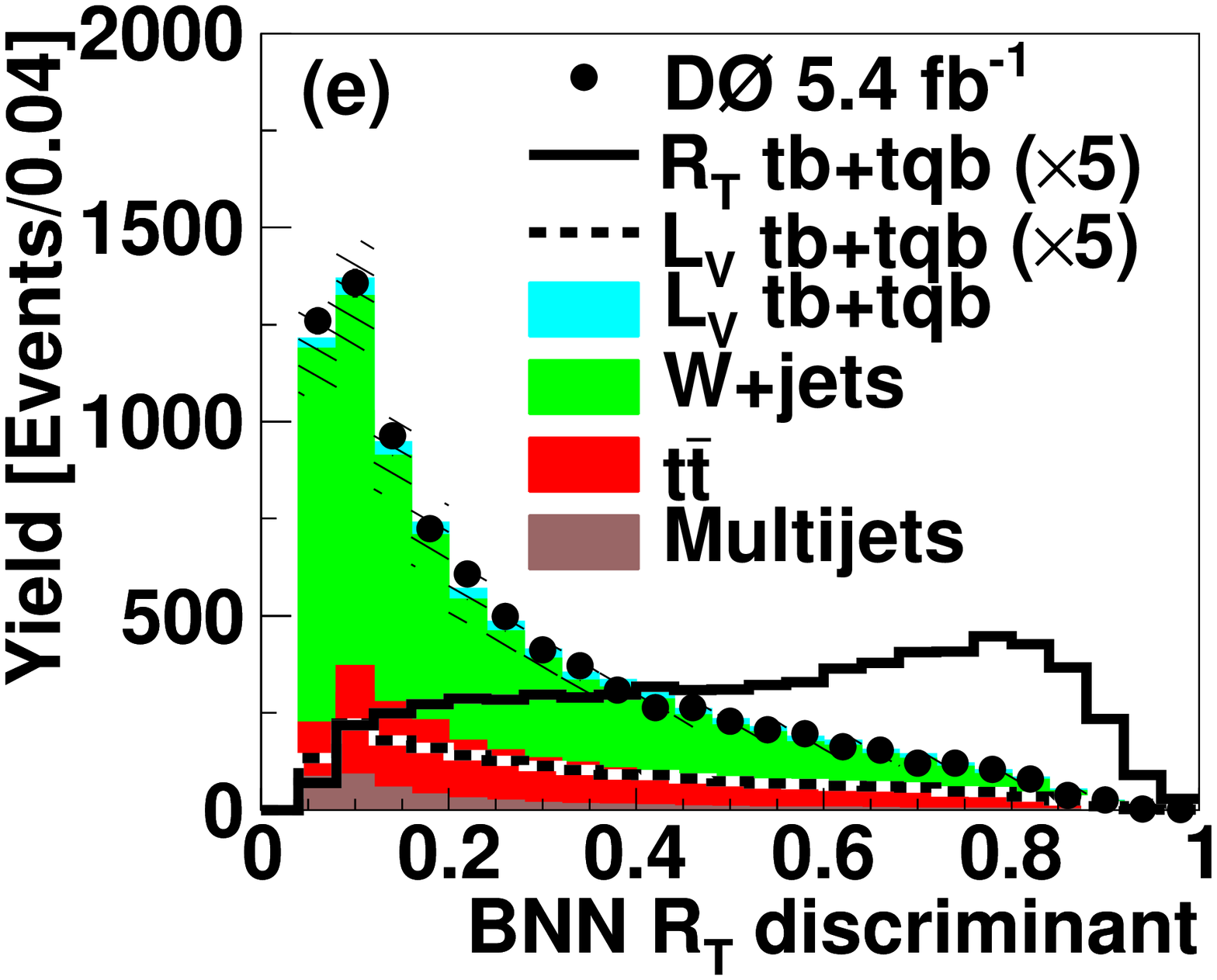}
\includegraphics[width=0.235\textwidth]{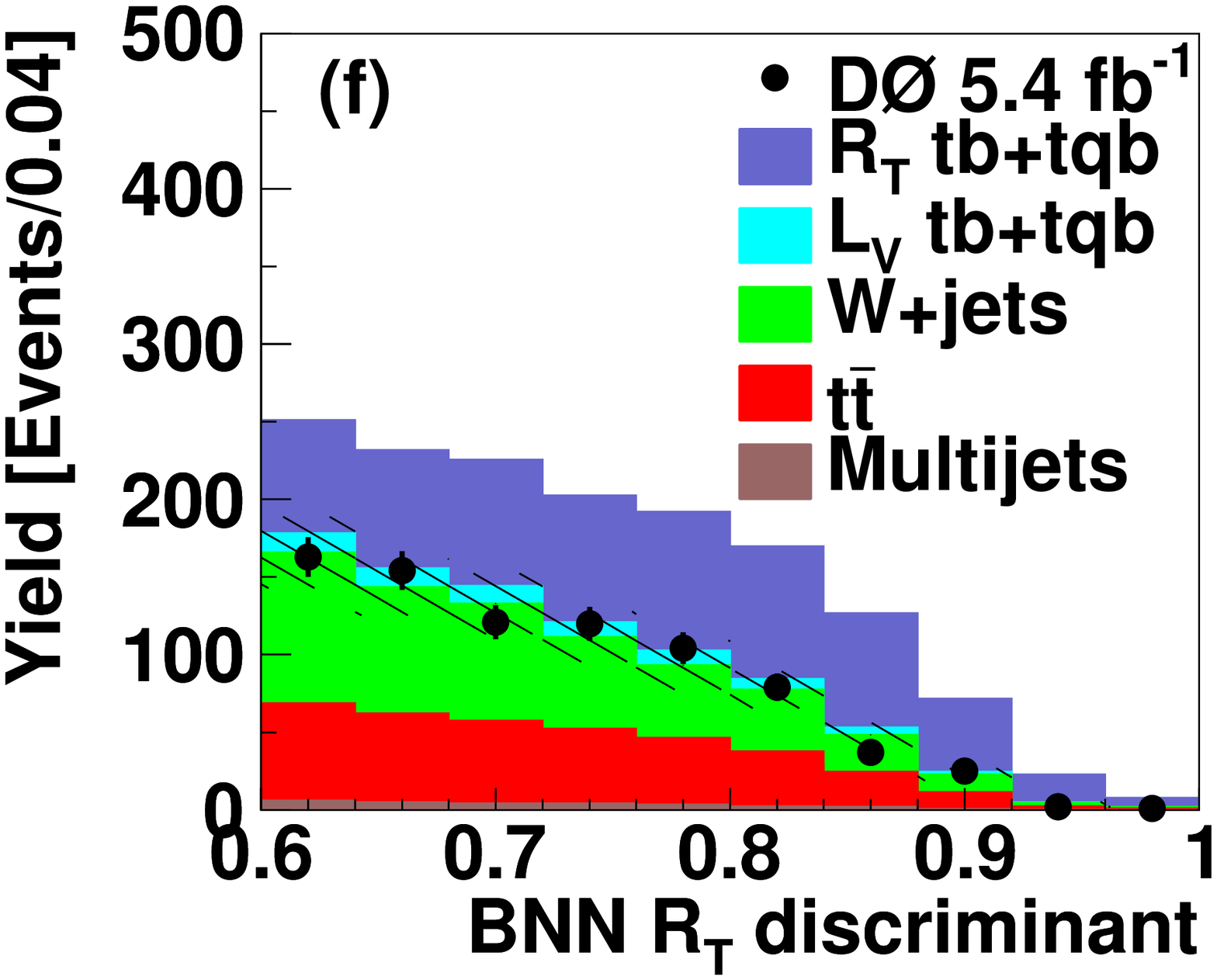}
\vspace{-0.1in}
\caption[mva_output]{The left column shows BNN discriminant output distributions of data and of the sum of the SM backgrounds with all 
channels combined for the whole discriminant range; superimposed are the distributions for the single top quark contributions scaled by 
(a) 5 times for the ($L_V$, $L_T$) scenario, (c) 20 times for the ($L_V$, $R_V$) scenario, and 
(e) 5 times for the ($L_V$, $R_T$) scenario. The right column shows BNN discriminant output distributions in the high discriminant 
region for (b) the ($L_V$, $L_T$) scenario, (d) the ($L_V$, $R_V$) scenario, and (f) the ($L_V$, $R_T$) scenario. 
The hatched bands give the uncertainty on the background sum. The $W$+jets contributions include the smaller
backgrounds from $Z$+jets and dibosons.
} 
\label{mva_output}
\vspace{-.1in}
\end{figure}

For each of the three scenarios, we consider the anomalous coupling sample as the 
signal when training BNN discriminants:  
for ($L_V$, $R_V$), the signal is the single top quark sample generated with $f_{R_V}=1$;
for ($L_V$, $R_T$), the signal is the sample generated with $f_{R_T}=1$;
for ($L_V$, $L_T$), the signal is the sample generated with $f_{L_T}=1$. 
The background includes the SM single top quark sample with $f_{L_V}=1$ and all the backgrounds described above.
Each background component is represented in proportion to its expected
fraction given by the background model.
Figure~\ref{mva_output} shows representative BNN discriminant output distributions for the three
different scenarios with all six analysis channels combined. 

We follow a Bayesian statistical approach~\cite{d0-prl-2007, d0-prd-2008,
stop-obs-2009-d0} to compare data to the signal predictions given by
different anomalous couplings using BNN discriminant output distributions.
We compute a two-dimensional (2D) posterior probability as a function of
$|V_{tb} \cdot f_{L_V}|^2$ and $|V_{tb} \cdot f_{X}|^2$,
where $V_{tb} \cdot f_{X}$ is one of the two non-SM couplings $X =
\{R_V, L_T\}$.
For these two cases the single top quark contribution is represented by
a superposition of two samples:
\begin{equation}
s = |V_{tb} \cdot f_{L_V}|^2 s_{L_V} +  |V_{tb} \cdot f_{X}|^2 s_{X},
\end{equation}
where $s_{L_V}$ ($s_{X}$) are the mean expected count of
single top quarks for the assumptions
$f_{L_V} = 1$ ($f_{X} = 1$) and the other couplings are set to zero.
In the $(L_V, R_T)$ scenario, the two couplings interfere, and to account for
the effect of the interference, the single top quark contribution is
represented by the superposition of three samples:
\begin{eqnarray}
s & = & |V_{tb} \cdot f_{L_V}|^2 s_{L_V} +  |V_{tb} \cdot f_{R_T}|^2 s_{R_T}
+\nonumber\\
 & + & |V_{tb} \cdot f_{L_V}| |V_{tb} \cdot f_{R_T}| (s_{L_VR_T} - s_{L_V} - s_{R_T} ),
\end{eqnarray}
where $s_{R_T} $ is the mean count assuming a left-handed tensor coupling only $f_{R_T}=1$,
and $s_{L_VR_T}$ is the one where both couplings $f_{L_V}=1$ and $f_{R_T}=1$.
The last sample is indicated as ``$L_V + R_T$" in Fig.~\ref{mva_output}f.
We assume a Poisson distribution for data counts and uniform prior probability
for nonnegative values of the SM and non-SM couplings.
The output discriminants for the signal, backgrounds, and data are used to
form a binned likelihood as a product over all six analysis channels and all bins,
taking into account all systematic uncertainties and their correlations.
The expected posterior probabilities are obtained by setting
the number of data counts to be equal to the predicted sum of the signal and backgrounds.

\begin{figure*}[!h!tbp]
\centering
\includegraphics[width=0.30\textwidth]{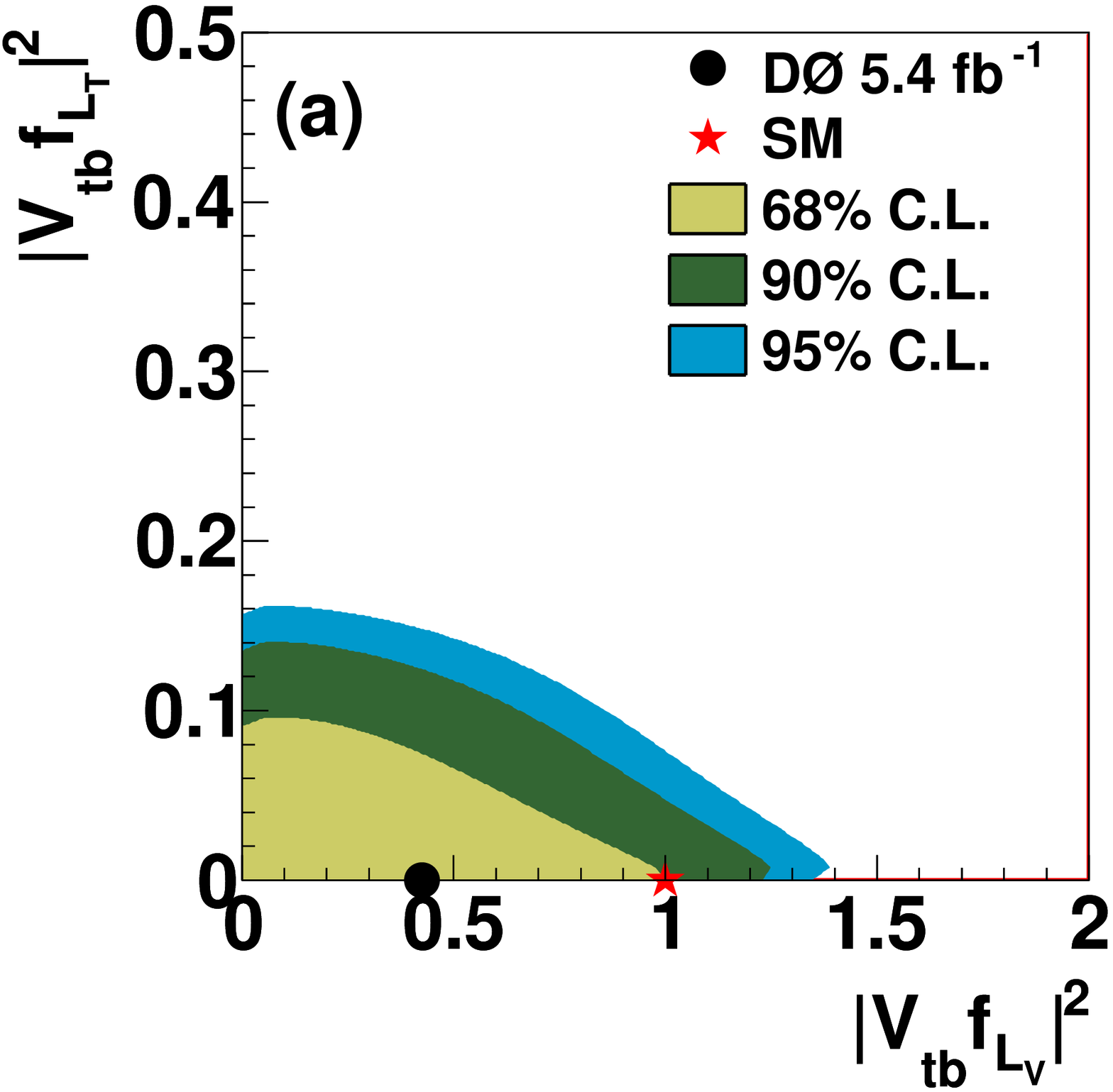}
\includegraphics[width=0.30\textwidth]{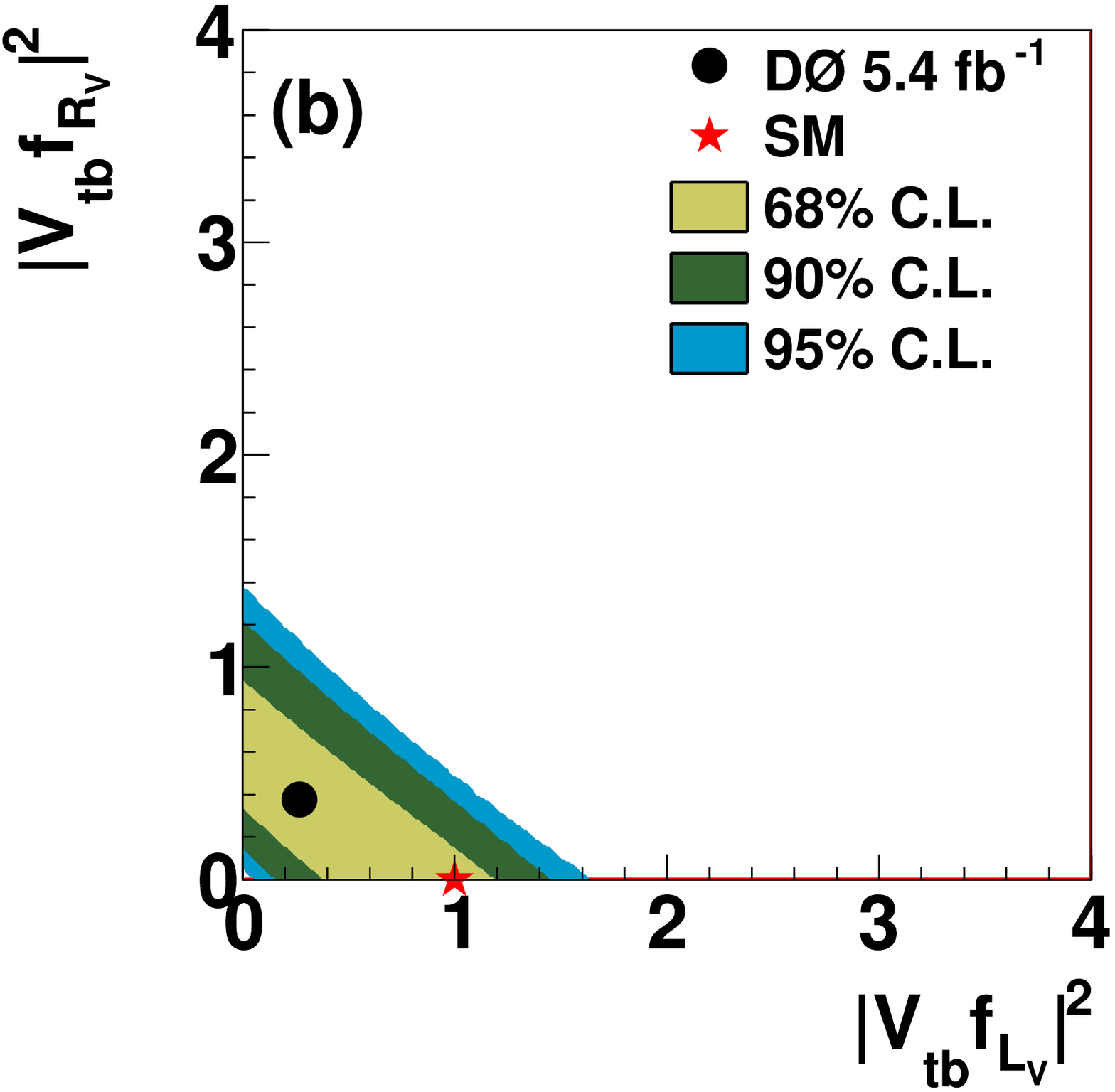}
\includegraphics[width=0.30\textwidth]{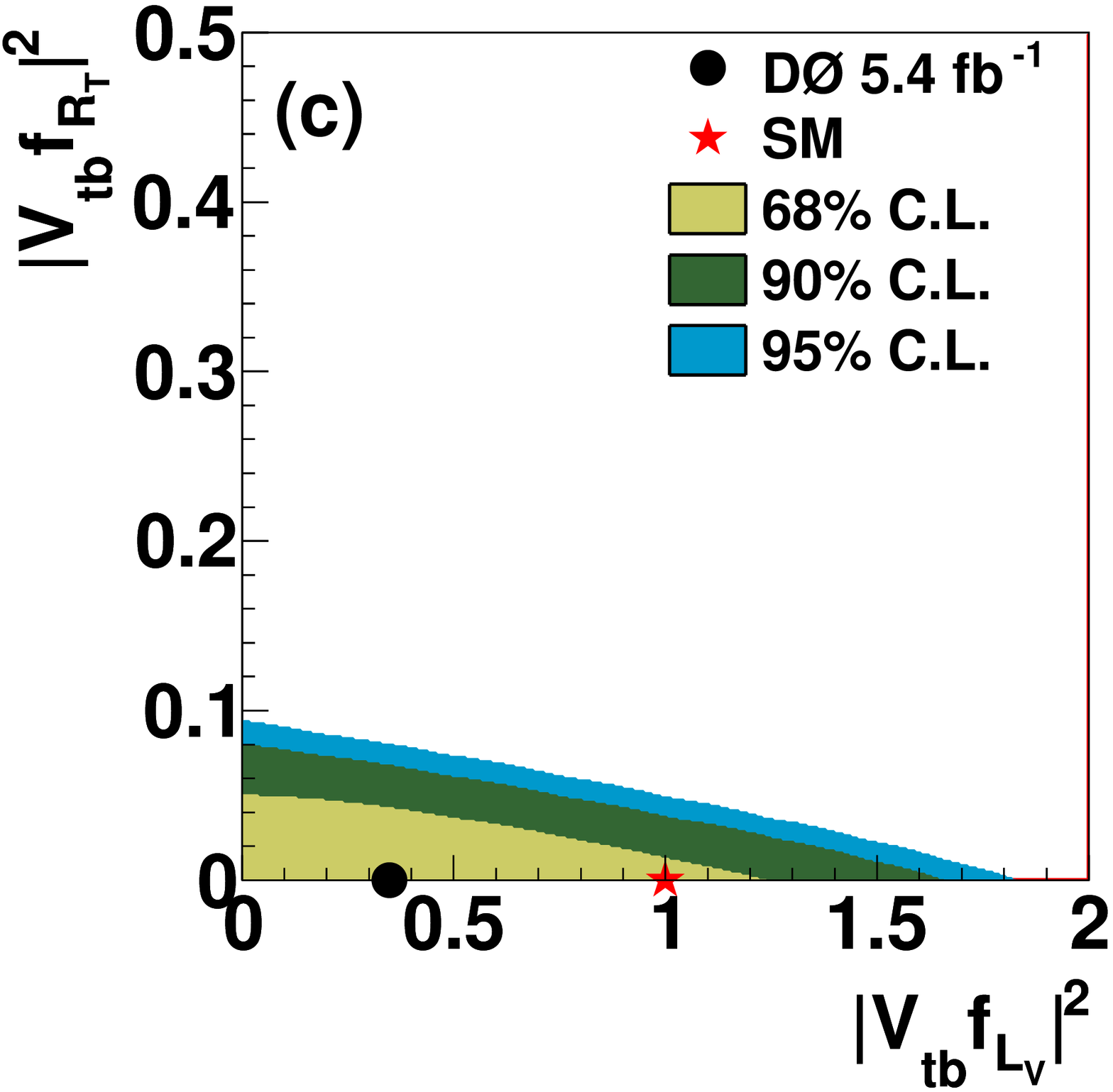}
\caption[measfullsys_2D]{Two-dimensional posterior probability density distributions for the anomalous couplings. 
The left row (a) shows the distribution for the ($L_V$, $L_T$) scenario, the middle row (b) for 
the ($L_V$, $R_V$) scenario, and the right row (c) for the ($L_V$, $R_T$) scenario. The dots represent the peak
posterior from our data in comparison with the SM predictions.}
\label{fig:measfullsys_2D}
\vspace{-.1in}
\end{figure*}

Figure~\ref{fig:measfullsys_2D} shows the 2D posterior probability density distributions for the three scenarios. 
We do not observe significant deviations from the SM expectations and therefore compute 
95\% C.L. upper limits on the anomalous couplings by integrating out the left-handed vector coupling to 
get a one-dimensional posterior probability density. The measured values are given in Table~\ref{table:obslim}.  
With the SM constraint on the left-handed vector coupling, i.e. 
$|V_{tb} \cdot f_{L_V}| = 1$, the 95\% C.L. limits on left-handed tensor, right-handed vector and 
tensor couplings are $|V_{tb} \cdot f_{L_T}|^2 < 0.05$, $|V_{tb} \cdot f_{R_V}|^2 < 0.50$ and 
$|V_{tb} \cdot f_{R_T}|^2 < 0.11$, respectively.

\begin{table}[t!!!]
\caption{\label{table:obslim} One-dimensional upper limits at 95\% C.L. for anomalous $Wtb$ couplings in the three scenarios.}
\begin{ruledtabular}
\begin{tabular}{lcl}
Scenario  & Cross section  &  \multicolumn{1}{c}{Coupling}   \\
\hline \\
$(L_V, L_T)$  & $<0.60$~pb 
               & $|V_{tb} \cdot f_{L_T}|^2<0.06$   \\
$(L_V, R_V)$  & $<2.81$~pb  
               & $|V_{tb} \cdot f_{R_V}|^2<0.93$  \\
$(L_V, R_T)$  & $<1.21$~pb  
             & $|V_{tb} \cdot f_{R_T}|^2<0.13$  \\
\\
\end{tabular}
\end{ruledtabular}
\vspace{-.1in}
\end{table}

In summary, we have presented a search for anomalous $Wtb$ couplings using 5.4~fb$^{-1}$ of D0~data in the 
single top quark final state. We find no evidence for anomalous couplings and set 95\% C.L. limits
on these couplings. These represent improvements in the limits by factors of 2.6 to 5.0 in terms of couplings squared 
compared to the previous results~\cite{anom-prl-2008} while a factor of approximately 2.5 is expected from the 
increase in integrated luminosity. 
This result represents the most stringent direct constraints on anomalous $Wtb$ interactions.

%
We thank the staffs at Fermilab and collaborating institutions,
and acknowledge support from the
DOE and NSF (USA);
CEA and CNRS/IN2P3 (France);
FASI, Rosatom and RFBR (Russia);
CNPq, FAPERJ, FAPESP and FUNDUNESP (Brazil);
DAE and DST (India);
Colciencias (Colombia);
CONACyT (Mexico);
KRF and KOSEF (Korea);
CONICET and UBACyT (Argentina);
FOM (The Netherlands);
STFC and the Royal Society (United Kingdom);
MSMT and GACR (Czech Republic);
CRC Program and NSERC (Canada);
BMBF and DFG (Germany);
SFI (Ireland);
The Swedish Research Council (Sweden);
and
CAS and CNSF (China).

\vspace{-0.3in}


\end{document}